\documentclass[12pt]{spieman}  % 12pt font required by SPIE;
\usepackage{amsmath,amsfonts,amssymb}
\usepackage{graphicx}
\usepackage{setspace}
\usepackage{tocloft}
\usepackage{lineno}
\usepackage{authblk}
\usepackage{gensymb}
\usepackage{soul,color,xcolor}
\usepackage{orcidlink}
\soulregister\cite7
\soulregister\ref7
\usepackage{hyperref}

\title{Design and characterization of W-band and D-band calibration sources for the AliCPT-1 Experiment}

\author[1]{Xu-Fang Li\orcidlink{0000-0002-2793-9857}}
\author[1,*]{Cong-Zhan Liu}
\author[1]{Ai-Mei Zhang}
\author[1]{Zheng-Wei Li}
\author[1]{Xue-Feng Lu}
\author[1]{Zhong-Xue Xin}
\author[1]{Guo-Feng Wang}
\author[1]{Yong-Ping Li}
\author[1]{Yong-Jie Zhang}
\author[1]{Shi-Bo Shu}
\author[1]{Yi-Fei Zhang}
\author[1]{Ya-Qiong Li}
\author[1]{Zhi Chang}
\author[1]{Dai-Kang Yan\orcidlink{0000-0003-4358-7959}}

\affil[1,*]{State Key Laboratory of Particle Astrophysics, Institute of High Energy Physics, Chinese Academy of Sciences, 19B Yuquan Road, Shijingshan District, Beijing, 100049, China\\}
\cftpagenumbersoff{figure}
\cftpagenumbersoff{table} 
\begin{document}
%\sethcolor{yellow}
\maketitle

\begin{abstract}

Ali CMB Polarization Telescope (AliCPT-1) is the first Chinese cosmic microwave background (CMB) experiment aiming to make sensitive polarization maps of the potential B-mode signal from inflationary gravitational waves. The telescope was deployed on Tibet Ali site at 5250 m above sea level in early 2025. Before and after each observation season, the instrument performance must be carefully calibrated, including the far field beam performance, far side lobe, spectral response, polarization angle and cross-polar beam response. To characterize these optical performances, several calibrators had been developed. We developed a W-band source and a D-band source for AliCPT-1 telescope's beam characterizations. In this paper, we present the design and performance of the two calibration sources.  

%This document shows the required format and appearance of a manuscript prepared for SPIE journals. It is prepared using LaTeX2e with the class file \texttt{spieman.cls}. Please note that the following journals require the use of structured abstracts in manuscript submissions: \textit{Biophotonics Discovery}, \textit{Neurophotonics}, the \textit{Journal of Biomedical Optics}, and the \textit{Journal of Medical Imaging}. Structured abstracts are encouraged for the \textit{Journal of Micro/Nanolithography, MEMS, and MOEMS}. Guidelines are available on the journal website. Whether structured or single-paragraph, the abstract should be a summary of the paper and not an introduction. Because the abstract may be used in abstracting and indexing databases, it should be self-contained (i.e., no numerical references) and substantive in nature, presenting concisely the objectives, methodology used, results obtained, and their significance. A list of up to six keywords should immediately follow. 
\end{abstract}

% Include a list of up to six keywords after the abstract
\keywords{AliCPT-1, cosmic microwave background polarization, Calibration, Beam}

% Include email contact information for corresponding author
{\noindent \footnotesize\textbf{*}Corresponding author name,  \linkable{liucz@ihep.ac.cn} }

\begin{spacing}{2}   % use double spacing for rest of manuscript

\section{Introduction}
\label{sect:intro}  % \label{} allows reference to this section
The Cosmic Microwave Background (CMB) was the earliest photon radiation in the Universe. It originated approximately 380,000 years after the Big Bang when photons decoupled from the electrons at the last scattering surface. It carries a wealth of information about the early Universe. Precise measurements of the CMB, especially the detection of the CMB \textit{B}-mode polarization patterns, are very important for understanding the origin and evolution of the universe. 

Ali CMB Polarization Telescope (AliCPT-1) \cite{Li2018} is a unique high altitude experiment for CMB observations in the Northern Hemisphere. It is at Ali astronomical observatory B1 point (80\degree 01' E,32\degree 18' N) with 5250m altitude in the Tibetan plateau in China. The main science goals of AliCPT-1 is probing the primordial CMB B-mode polarization pattern. AliCPT-1 is designed as a refracting telescope with a medium-size aperture (72 cm), resulting in angular resolution 19' and 11' at 90 and 150 GHz, respectively. In the early 2025, the phase-I of AliCPT-1 was deployed at the site with one dichroic detector module consisting of 432 pixels, which is fabricated by NIST. Each pixel has four transition-edge sensors (TES) to detect the orthogonal polarizations in two separate frequency bands (90 GHz and 150 GHz) \cite{Salatino2020,Salatino2021}. Any beam shape mismatch between the co-located detector pair will induce temperature-to-polarization leakage and cause spurious \textit{B}-mode signals. So the instrument optical performance must be precisely characterized, including the band-pass response, polarization, beam mapping and far sidelobes etc. For calibrating the beam characterization of AliCPT-1, two polarized sources in W-band and D-band had been developed in the lab. In this paper we present the instrument design and performance evaluation of the two calibration sources. 

In section \ref{sec2}, we describe the beam measurement setup , requirement for the sources and the source design and construction.  In section \ref{sec3}, we present the instrument performance tested in lab. At last, the conclusions are provided in section \ref{sec4}.

%This document shows the format and appearance of a manuscript prepared for submission to SPIE journals. Note that this template is only intended to be used as a guideline for author convenience. It is designed for optimum clarity and ease of reading for editors and reviewers, but the template does not reflect the final page layout of a published journal paper. Accepted papers are professionally typeset in XML according to the layout and design of the journal. 

\section{Millimeter-wave source design}\label{sec2}

\subsection{Far field beam measurement setup and requirement for calibration source} \label{sec2.1}
\begin{figure}[H]
\centering
\includegraphics[width=0.8\textwidth]{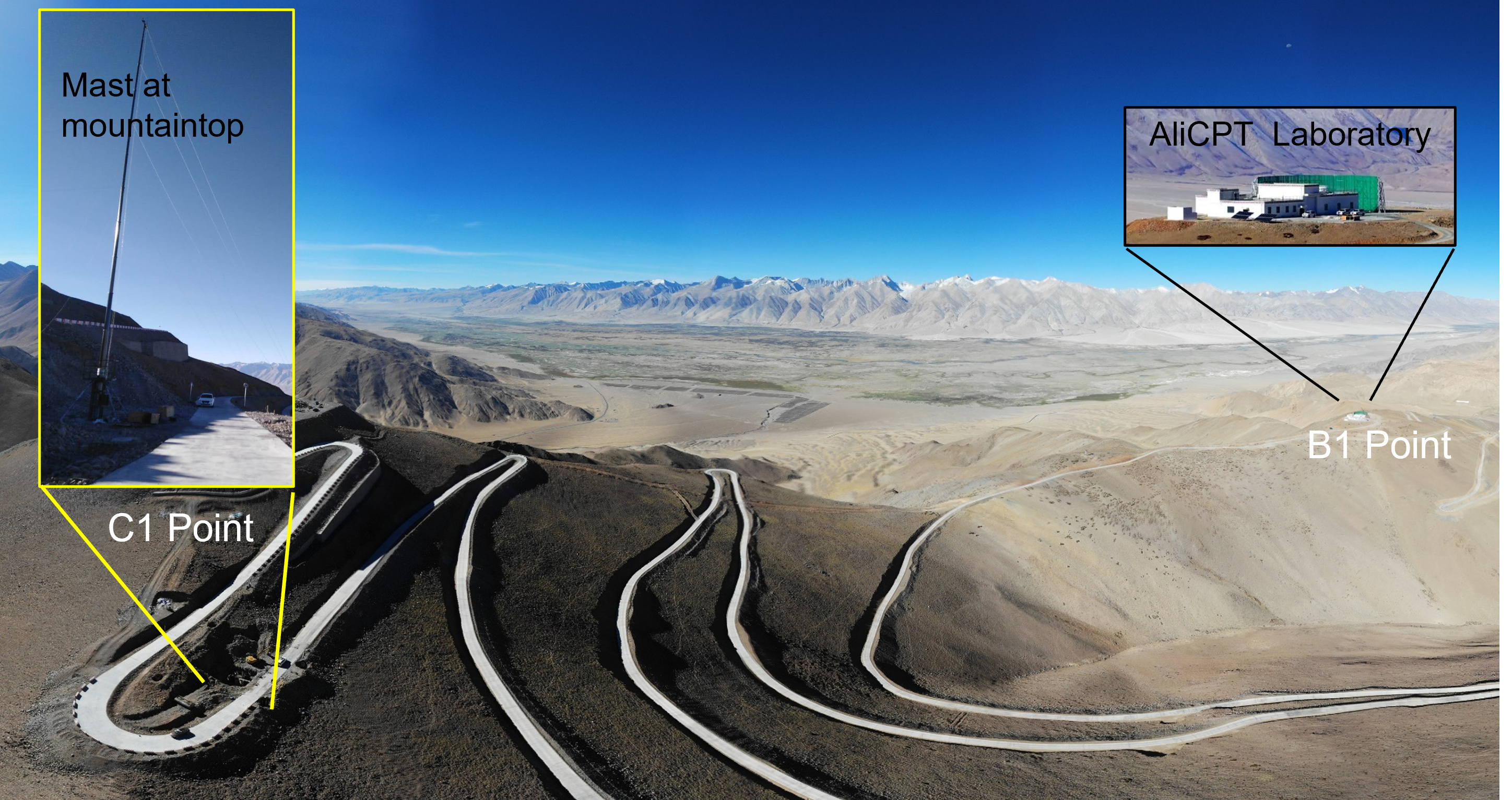}
\caption{\label{fig：siteandffbmloc}The source would be mounted on a 28 m tall mast at Ali astronomical observatory C1 point, which is 1400 m far away from B1 point where AliCPT laboratory is. }
\end{figure}

According to the far-field condition \textit{$2D^2/\lambda$}, the distance between the calibration sources and the AliCPT-1 telescope should be greater than 311/518 m for 90/150 GHz, respectively. Based on the terrain around AliCPT-1 site, the C1 point hilltop, which is 1,400 m away and 170 m higher, is a suitable place to install the calibrator, showed in Fig \ref{fig：siteandffbmloc}. To mitigate the pick-up signal of C1 hilltop into the main beam of AliCPT-1, a 28m pneumatic mast was equipped to mount the calibration sources atop. Since the telescope can only tip down to $\sim 45\degree$ from zenith, a 2 m $\times$ 3 m ultra-flat low-deformation aluminum mirror \cite{Cai2024} was constructed to redirect the beams towards the far-field beam map (FFBM for short) calibrator. BICEP/Keck telescope located in South Pole used a mechanically chopped thermal source atop a mast beyond the far-field distance to perform the far field beam mapping \cite{Karkare2016,Giannakopoulos2024,Germine2020}, and the observed optical power from the thermal source was several $pW$. For AliCPT-1, the FFBM calibrator is much further than that of BICEP/Keck due to the larger aperture. A chopped blackbody thermal source with a radius of 1.7 m would need to be constructed to achieve the required several $pW$ power on the AliCPT-1 detectors if we were to adopt the scheme similar to the BICEP/Keck, that is too big to realize in practice. Given this impracticality, we turned to an alternative approach by using a microwave source. This method has been successfully demonstrated in other CMB experiments to characterize detector's polarization properties\cite{Buder2014,Cornelison2022,Nati2017,Ade2025,Coppi2025fmt}. Consequently, we have chosen to construct our FFBM calibrator using a microwave source for AliCPT-1, since it is almost the only choice due to its big aperture. 

In addition to the far-field beam of the telescope, we also need to use the source to evaluate the far sidelobe level. In the far sidelobes measurement (FSLM for short) campaign, the source is mounted on a mast 12m from the telescope, at $\sim 49\degree$ elevation and 18m away, so the far sidelobes can be mapped without flat mirror. 

Based on the Friis transmission equation \cite{friis_formula} (Equation \ref{eq:Friis}), we calculated the needed source power and the dynamic range.  
\begin{equation}
   \ P_r = (\emph{PLF}) \cdot P_t \cdot G_t \cdot G_r \cdot (\frac{c}{4 \pi \nu D})^2 \label{eq:Friis}
\end{equation}
Where $P_r$ is the power received at the receive antenna in Watts, $P_t$ is the power transmitted from the transmitting antenna in Watts, $G_t$ is the gain of the transmit antenna in the direction of receive antenna, $G_r$ is the gain of the receive antenna, $\nu$ is the frequency of the signal in Hz, $D$ is the distance between the transmitting and receiving antennas in meters, $c$ is the speed of light, \emph{PLF}  is Polarization Loss Factor if the antennas are not polarization matched, and we set it to 1/2 to account for the source illuminate the orthogonal detector pair simultaneously and equally. 
The equation can also be expressed in decibels:
\begin{equation}
 \ P_{r(dB)}=P_{t(dB)}+G_{r(dB)}+G_{t(dB)}-Los-3\ 
 \label{eq:FriisinDb}
\end{equation}

Where \emph{Los} is the path loss, \[Los=92.44+20lg D_{(km)} +20lg f_{(GHz)}.\]

 For AliCPT-1 detector module, each detector contains four polarization-sensitive AlMn TES covering two frequency bands for sky observations. In series to each AlMn TES, there is a pure Al TES with a high saturation power to allows for operation in high-load situations. We use the pure Al TES to conduct the FFBM calibration. The saturation power ($P_{sat}$) of Al TES is up to about 400 $pW$, while the detection sensitivity is less than 1 $fW$. So the source power received by the detector ($P_r$) should be more than ten times sensitivity to several $fW$ and less than $P_{sat}$ during calibration campaign. Both the gain of the two source horns ($G_t$) are nearly 22 dBi.
The gain of AliCPT horn is designed as 56 dBi at 90GHz and 60.4 dBi at 150GHz respectively. In the FFBM calibration campaign the beam profile should cover $(0 \sim -35)$ $ dB$. And in the FSLM campaign, the beam profile should reach to -90 dB. D is 1.4 km in FFBM campaign and 18 m in FSLM campaign.

Based on these parameters, the required output power for the FFBM and FSLM campaigns were calculated. For FFBM calibration, the W-band RF source must have an output power (Pt) between -16.1 dBm and 0.2 dBm, and the D-band RF source must have an output power (Pt) between -16.8 dBm and -1.2 dBm. During FSLM testing that covers the radiation pattern from 0 dB to -90 dB, the RF source must deliver more than -3 dBm at the -90 dB point and less than -42 dBm at the 0 dB point. With an additional margin of several dB, the required output power of the source ($P_t$) was decided to exceed 10 dBm (10 mW) with a dynamic range of 60 dB. 

Since the source is linearly polarized and one hexagonal detector module contains three sets of detector pairs with different polarization orientations, the orientation angle of the calibrator source should be adjustable when calibrating FFBMs of different pixels.

In order to improve the signal-to-noise ratio, to reduce the background interference and to enhance testing sensitivity, it is imperative to employ a chopped source. 

The AliCPT-1 is designed to operate with a broad bandwidth of 38GHz at 90GHz band and 40GHz at 150GHz. Accordingly,  the FFBM calibrator should function as a broadband source. For a continuous-wave source, its output power varies with frequency. To ensure the received power to be stable, the full band sweep time must be significantly shorter than the thermal time constant of the TES detector, which is generally several hundreds of $\mu s$ to several milliseconds. Simulations show that a sweep time of 8 $\mu s$ is sufficient to keep the TES output power stable for most of the applications. For effective data demodulation, the calibration source requires an integrated time synchronization capability.

\subsection{Instrument design}\label{sec2.2}
\begin{figure}[H]
\centering
\includegraphics[width=0.8\textwidth]{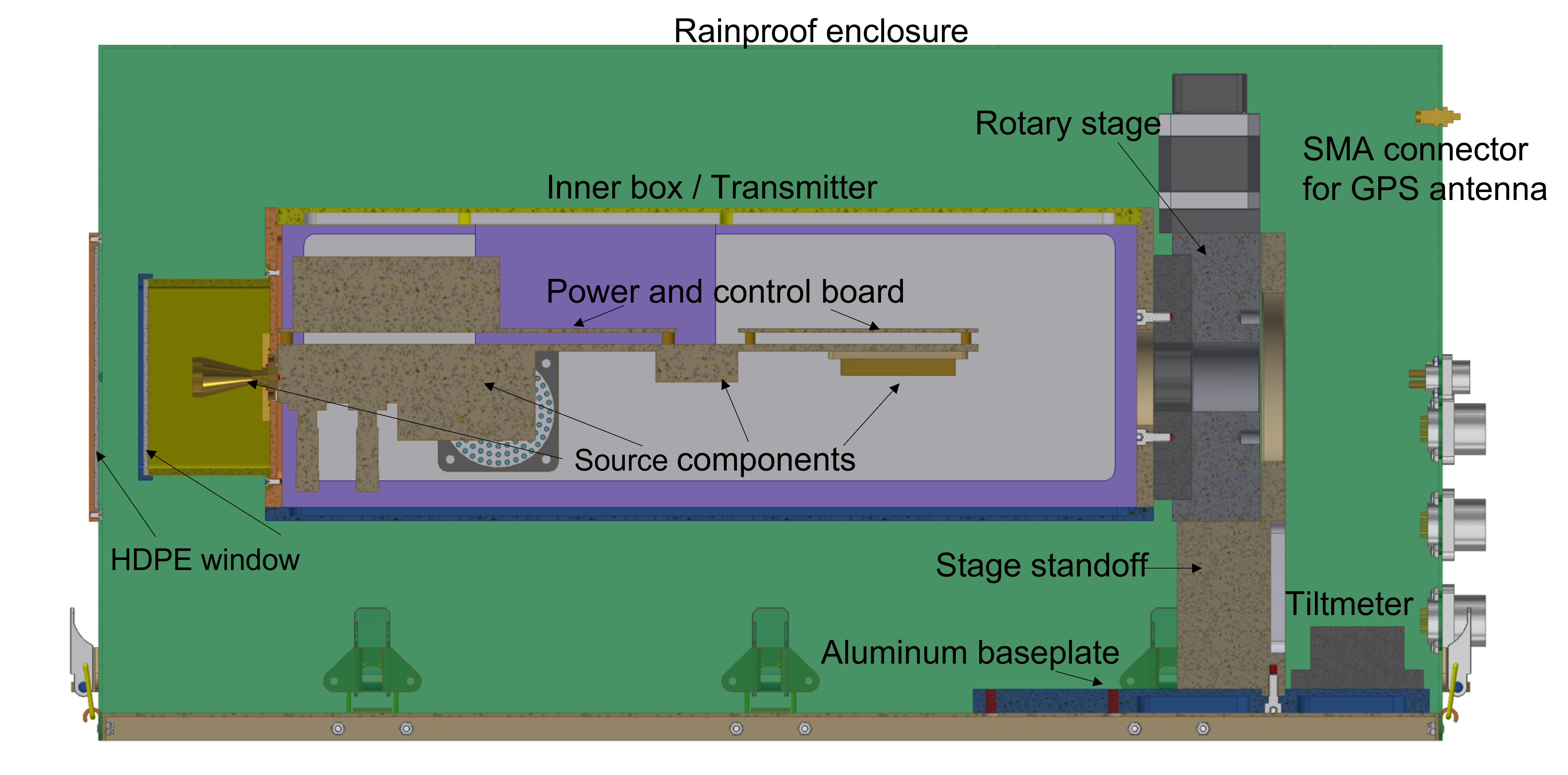}
\caption{\label{fig:schematic}Schematic of the W-band source for AliCPT-1 FFBM/FSLM calibration. The transmitter, which is enclosed in the inner box, contains the source components, power supply and control system. It is the core functional unit of the instrument that emits 75-110 GHz linear polarization radiation. The transmitter is mounted to a rotation stage so the angle of the polarization can be rotated in a range of $0  \sim 360\degree$. The rotation stage with the transmitter is bolted to an L-shaped standoff, which is rigidly bolted to an aluminum base plate. A tiltmeter is also bolted to the same base plate, so that the orientation of the polarization radiation can be monitored. The base plate will be mounted to the platform atop the mast.} 
\end{figure}

\begin{figure}[H]
\centering
\includegraphics[width=0.8\textwidth]{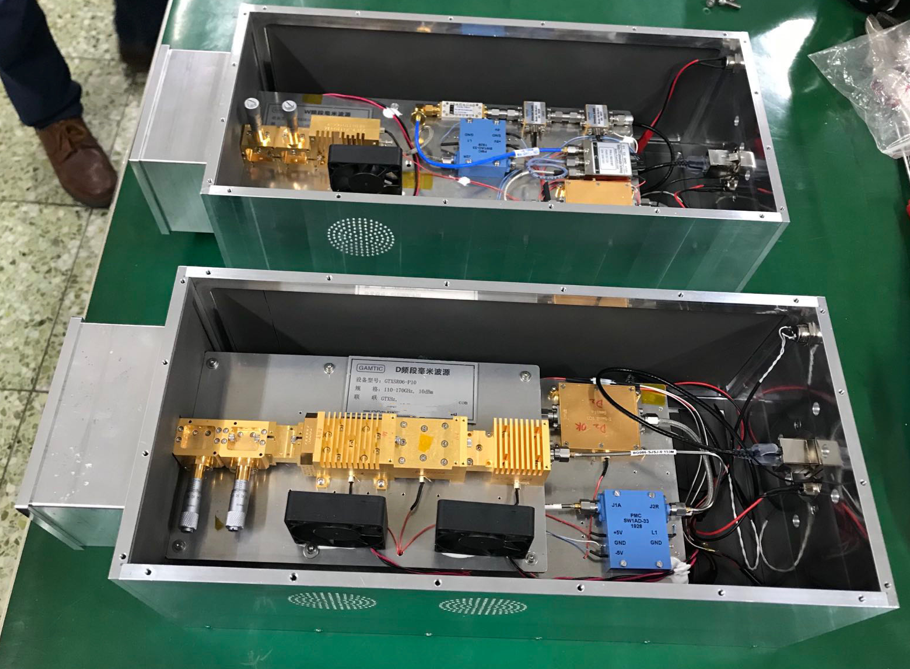}
\caption{\label{fig:transmiter} The W-band and D-band transmitters photoed before the inner box was covered.}
\end{figure}

In order to carry out the FFBM/FSLM calibration for AliCPT-1, two instruments that meet the functional requirements mentioned previously have been developed. Figure \ref{fig:schematic} shows the schematic of the W-band millimeter-wave source. The calibrator includes a transmitter that emits modulated wave of designed frequency, a rotary stage and its mechanical support, a tiltmeter, a base plate and a rainproof enclosure. The D-band millimeter-wave source has a nearly identical design, except that its transmitter uses different components  (e.g. multipliers, amplifiers, or antennas)  optimized for the D-band frequencies and has a modified structure. Figure \ref{fig:transmiter} shows the W-band and D-band transmitters. 

\begin{figure}[H]
\centering\includegraphics[width=0.8\textwidth]{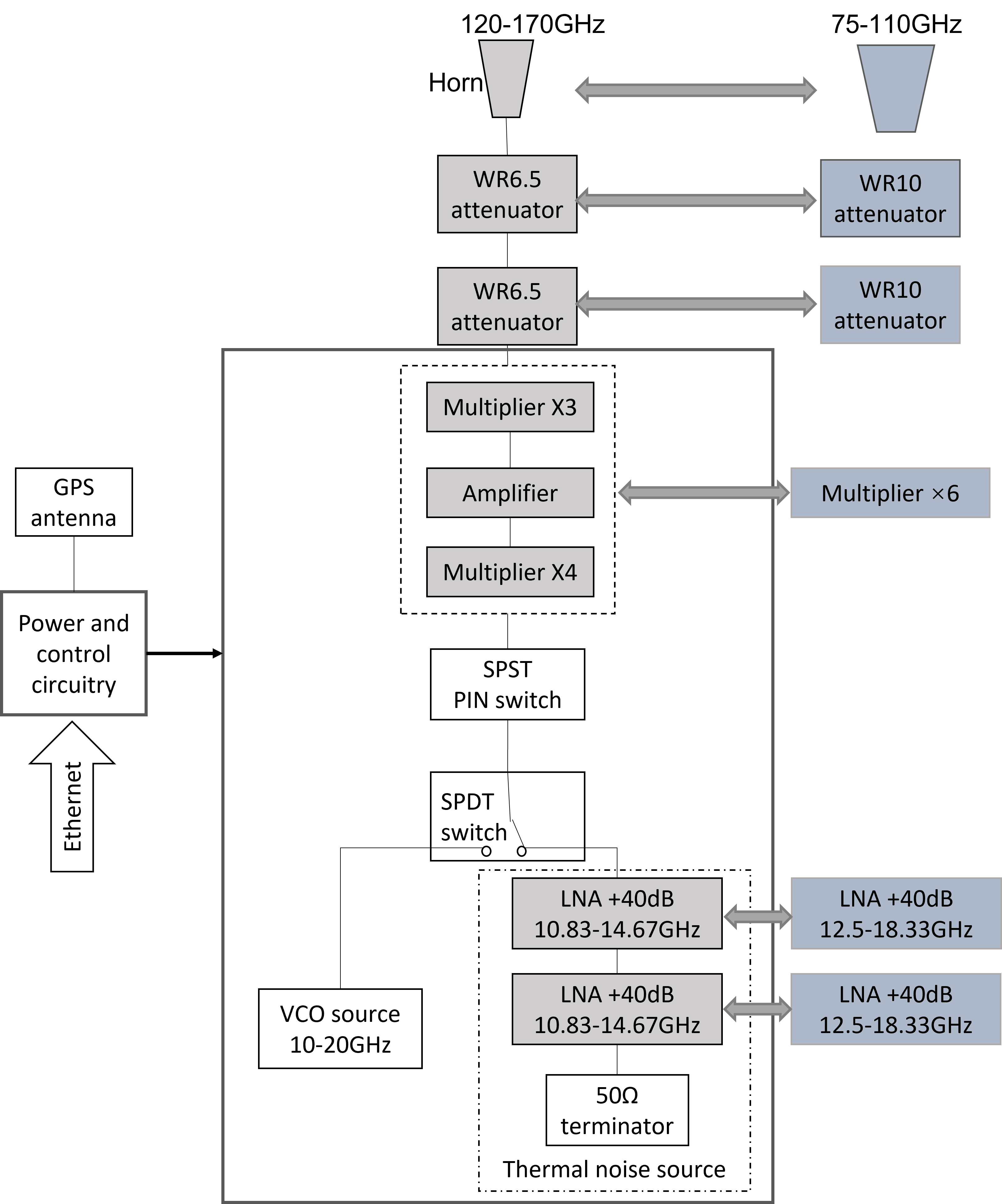}
\caption{\label{fig:ElectricalDiagram} Electrical Diagram of the W-band (75-110GHz) and D-band (120-170GHz) transmitters.  The microwave components of D-band transmitter are shown in gray and those of W-band transmitter are shown in blue gray.  }
\end{figure}

The transmitters are the core functional units, and their electrical diagrams are depicted in Figure  \ref{fig:ElectricalDiagram}. The source achieves the desired frequency band by multiplying and amplifying the fundamental frequency signal. There are two different channels to generate fundamental frequency radiation, which can be selectable via a Beijing CTE L1S21218T Single-Pole Double-Throw (SPDT) switch designed functionally for DC to 18GHz. One channel is the amplified thermal noise source, where two Spacek Labs low noise amplifier (LNA, SG156-40-2 amplifiers for W-band source, which are optimized for amplification of 12.5 to 18.33GHz, and SG134-40-17 amplifiers for D-band source, which are optimized for amplification of 10.83 to 14.67 GHz)  boost the blackbody spectrum emitted by the 50$\Omega$ resistor. The other channel is Gamtic VCO-1020KF voltage-controlled oscillator (VCO), which can be configured to operate at single-frequency mode or sweeping-frequency mode. A Pulsar SW1AD-33 PIN diode switch is used to chop the signal before it passes through the following  multipliers. The multipliers and amplifier (RPG AFM6 75-110+14 sextupler for W-band source, and AMC4-19-P15BPF U-band X4 multiplier, HPA-19-P23FB U-band high power amplifier and RPG FM3 110-170+10 multiplier for D-band source) are carefully selected to ensure that the output power exceeds 10 dBm  in the desired frequency range. Two RPG WTA 75-110 or WTA 110-170 waveguide tunable attenuators (WTA) allow the output power to be controlled over a range of more than 60dB. Finally the radiation is emitted out from a Custom Microwave (RCHO10R or RCHO6R) horn antenna. The power and control system comprises a power supply module, a STM32F103VBT6 MCU, a MAXII EPM240T100I5N CPLD, a high-speed DAC904, an analog signal conditioning circuit, and a GPS receiver. The MCU communicates with the computer via Ethernet, generating control signals for the SPDT and SPST, transmitting the VCO's operating mode and parameters to the CPLD, and recording the UTC time synchronized with the GPS pulse per second (PPS). Based on the received information, the CPLD controls the DAC to generate an analog waveform, which drives the VCO to produce the desired frequency signal. To match the time constant of the instrument detectors, the source operates in a continuous, adaptive high-speed sweep mode. Rather than using a fixed sweep pattern, the system dynamically optimizes sweep parameters based on user-defined start/stop frequencies and sweep period. The MCU first converts these into target DAC values and the total time in units of the system's minimum step duration (20 ns). An enumeration algorithm then searches for the optimal step configuration within the hardware constraint of a 20 ns minimum step duration and a performance requirement of typically exceeding 160 total frequency steps per sweep. The algorithm evaluates possible step durations (in multiples of 20 ns) to find the setting that minimizes the cumulative DAC error when distributing the required frequency change. Finally, the CPLD executes back-to-back sweeps using these optimized parameters, typically achieving sweep rates up to 125,000 sweeps per second (for an 8 $\mu s$ cycle) or as configured. This design ensures rapid, flexible, and accurate coverage of the desired frequency band. The MCU also records the sub-second time code based on its internal clock corresponding to each flip of the SPST. As a result, the source  modulating  time can be reestablished for the detector calibration data demodulation. The transmitter was enclosed in an aluminum inner-box with inner surface covered by Eccosorb  BSR microwave absorber,  and the aluminum surrounding the horn antenna was covered by Eccosorb HR-10 absorber.  

As the microwave components within the source transmit signals via standard rectangular waveguides, the emitted radiation from the source is linearly polarized. So we mounted the transmitter to a high precision rotation stage (Zolix TBRK100) to adjust its polarization orientation. There is an encoder on the stage so the position of the stage at any time can be reliably reported, and the semi-closed-loop resolution of the stage is 0.0005\degree,  and the repeat position accuracy is $\leq \pm 0.003\degree $ . The rotation support standoff and a tiltmeter are bolted to an aluminum base plate, which can be mounted to the platform atop the mast, so that the change in the orientation of the source on the mast can be monitored. 

\section{Instrument Performance}\label{sec3}

Tests were conducted throughout the construction process and after the integration to verify the full functionality of the two millimeter-wave polarized sources. The results showed that the calibrator can meet the requirements. 

\subsection{Power and dynamic range}\label{subsec3.1}

\begin{figure}[H]
\centering
\includegraphics[width=0.8\textwidth]{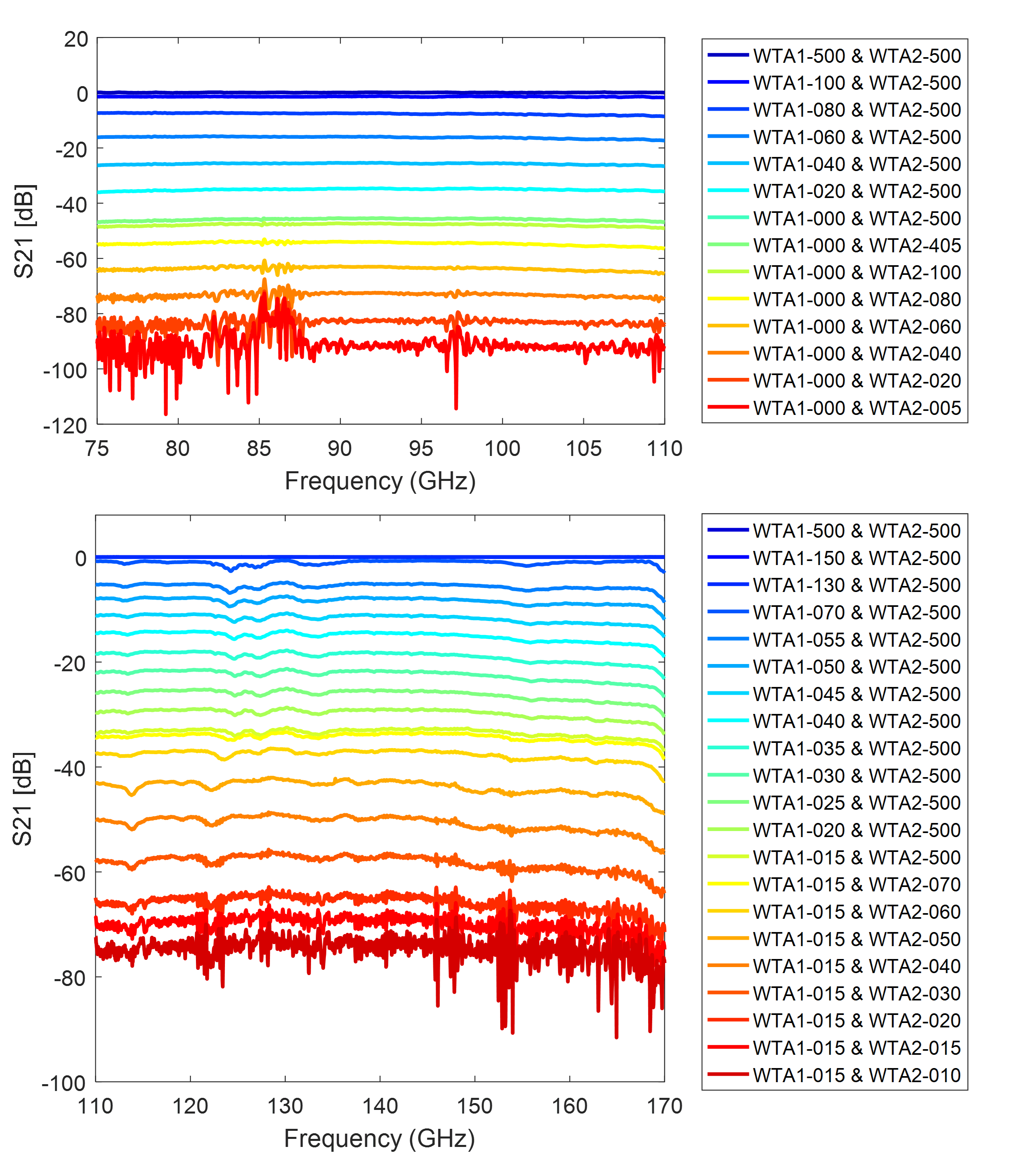}
\caption{\label{fig:VA_S21}  The S21 data for the two-stage WTAs (W-band: top; D-band: bottom) are normalized to their mechanical zero (micrometers fully retracted at WTA1-500, WTA2-500) to establish a consistent 0 dB reference, accurately showing the relative attenuation. WTA1 (input side, connected to frequency multiplier or Port 1 of VNA ) and WTA2 (output side, connected to antenna or  Port 2 of VNA) are labeled in the legend with suffixes indicating their micrometer readings.}
\end{figure}

The output power of the sources were tested with Keysight E4416A Power Meter, W8486A waveguide power sensor (75-110 GHz, -30 dBm to +20 dBm), and Elimka WTT-03E/02E Waveguide tapered transition WR-10 to WR-6.5 (for D-band source testing). During the test, the horn antenna of the source was removed, and the waveguide power sensor was directly connected to the waveguide of WTA. The source achieves a maximum output power of +14 dBm  (25.1mW)  and +11 dBm  (12.6mW) for the W-band source and D-band source at both thermal noise channel and VCO source channel.  The background noise level of this testing system is approximately -25 dBm in the W-band and -17 dBm in the D-band, and the designed sources dynamic range is greater than 60 dB. Consequently, using this testing system, the output power across the full range of the two WTAs’ micrometer positions could not be calibrated.   Instead,  we tested the S21 parameter of the two-stage WTAs with a Vector Network Analyzer(VNA).  The WTA connected to the multiplier is designated as WTA1, while the one interfacing with the output antenna is designated as WTA2. During testing, WTA1 was connected to the Port 1 of the VNA, and WTA2 was connected to the Port 2 of the VNA.  The S21 performance of the two WTAs  under different attenuation settings are shown in Figure \ref{fig:VA_S21}. The data are plotted as the relative response after normalization to the measurement taken at the mechanical zero position(micrometer fully retracted) of each attenuator. The results demonstrated that the achieved dynamic range could fully meet the design requirements for FFMB and FSLM calibration.

\subsection{Stability}\label{subsec3.2}

FFBM campaign is designed to be carried out by rastering all of the detectors across the calibration sources, and this campaign would be repeated several times with the telescope at several boresight angles. Given this operational procedure, power stability is a crucial performance metric. It was evaluated using two RF waveguide detectors. Their output amplitude was acquired by an NI USB-6366 data acquisition module, with a sampling rate configured according to the operational mode: 1kHz for the thermal noise channel or VCO single-frequency mode, and 2MHz for the VCO sweep mode. 

These stability measurement was performed with both sources at their maximum output power (achieved by full retracting the WTA micrometers). For the VCO-based source, this entailed operating in a continuous sweep mode. Each frequency sweep (80-108.8 GHz for W-band or 130-161 GHz for D-band) consisted of 200 steps and was completed in a period of 8$\mu s$, with zero interval between successive sweeps. This configuration results in a dwell time of approximately 40 ns per frequency point, ensuring rapid coverage of the entire band. 

\begin{figure}[H]
\centering
\includegraphics[width=0.8\textwidth]{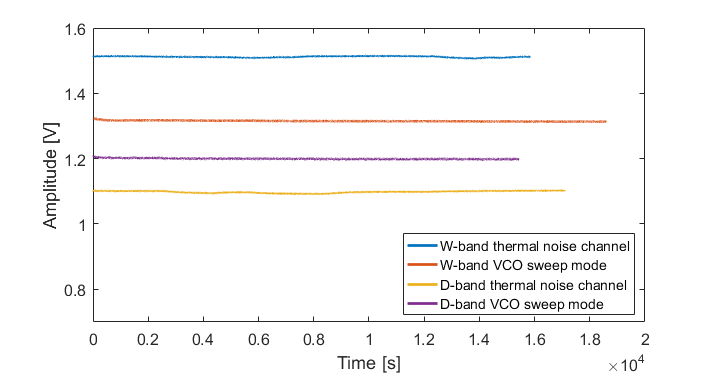}
\caption{\label{fig:stability}Output Power Amplitude (V) as a Function of Time(s) for both the thermal noise channel and the VCO sweep mode. The amplitude fluctuation remains within $\pm1\%$ over the 4-hours duration.}
\end{figure}

The average amplitude per second was recorded, and the stability was quantified by $(\textit{max}-\textit{min})/\textit{mean}$ of these one-second averages over the measurement period. Figure \ref{fig:stability} shows the corresponding time-sequence plot of the output power amplitude for both the thermal noise channel and VCO sweep mode, confirming the system’s stability.

When the sources worked at thermal noise channel, the stability over 4 hours was 0.52\% for W-band source and 0.97\% for D-band source. When  switched to the VCO-based sweep mode, the measured stability   over four hours was 0.3\% for W-band source and 0.5\% for D-band source. However, when the sources  operated in  single-frequency mode  via  the VCO channel, the output power  exhibited non-monotonic changes over time. Furthermore, both the trend and rate of these changes varied inconsistently across different tests. Thus, combined with our primary requirement to characterize the beam under broadband illumination—matching the actual observing conditions of our detectors—the single-frequency mode is unsuitable for far-field beam mapping.

\subsection{Spectra}\label{subsec3.3}

\begin{figure}[H]
\centering
\includegraphics[width = 0.9\textwidth]{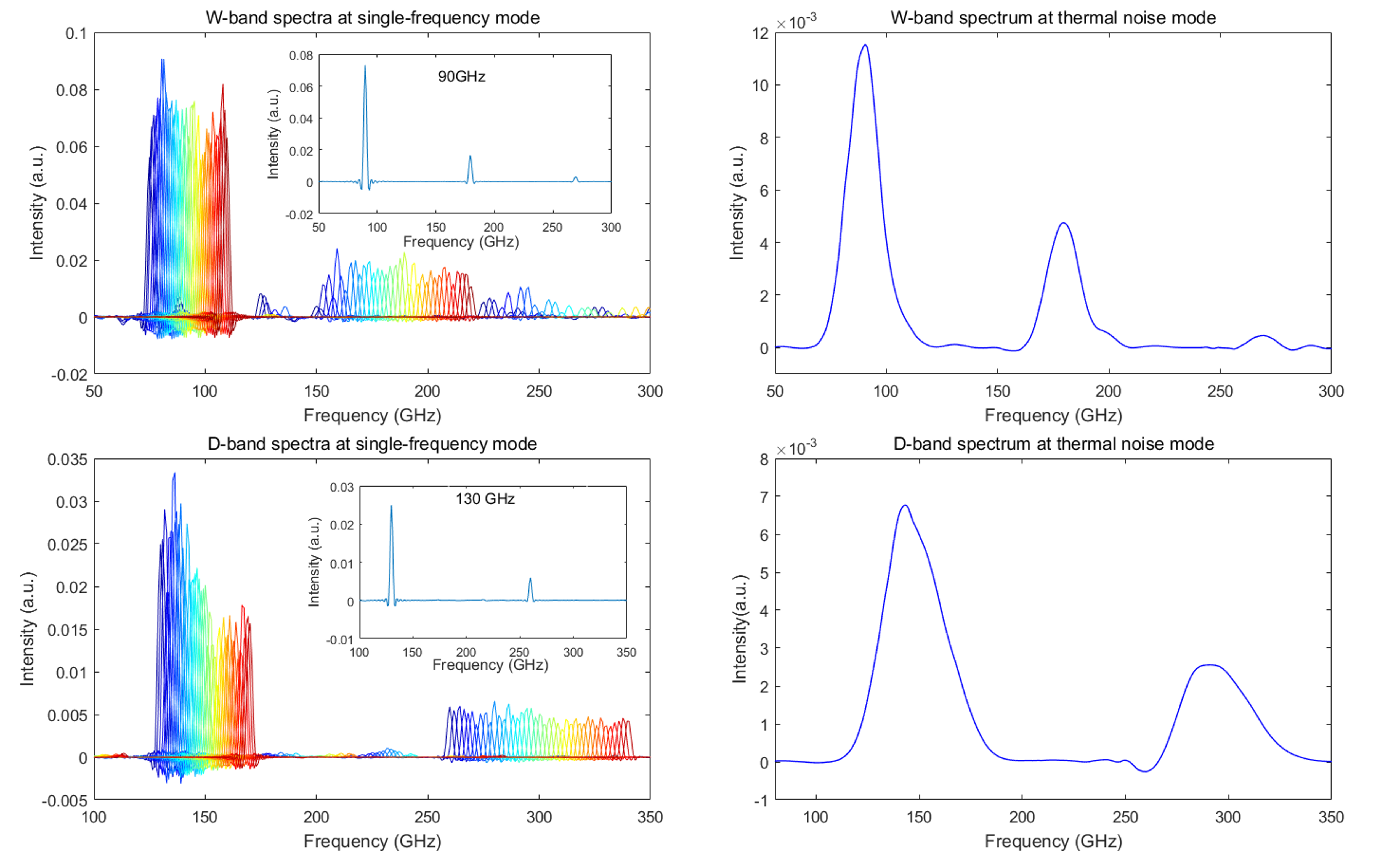}
\caption{\label{fig:spectra_tested_at_EAST} Spectra of the two sources tested with a rapid continuous-scan  FTS. The left two pannels are the monochromatic light spectra ranges from 75-110 GHz (upper left) and 130-170GHz (lower left)
when the source worked at single-frequency mode with a frequency step of 1GHz, and each spectrum is shown in different colors. The spectra of 90GHz and 130GHz are shown in separately to show details. The right two pannels are the spectra of the source worked at thermal noise channel. The harmonic components outside designed band is caused by sampling errors.  }
\end{figure}

\begin{figure}[H]
\centering
\includegraphics[width = 0.9\textwidth]{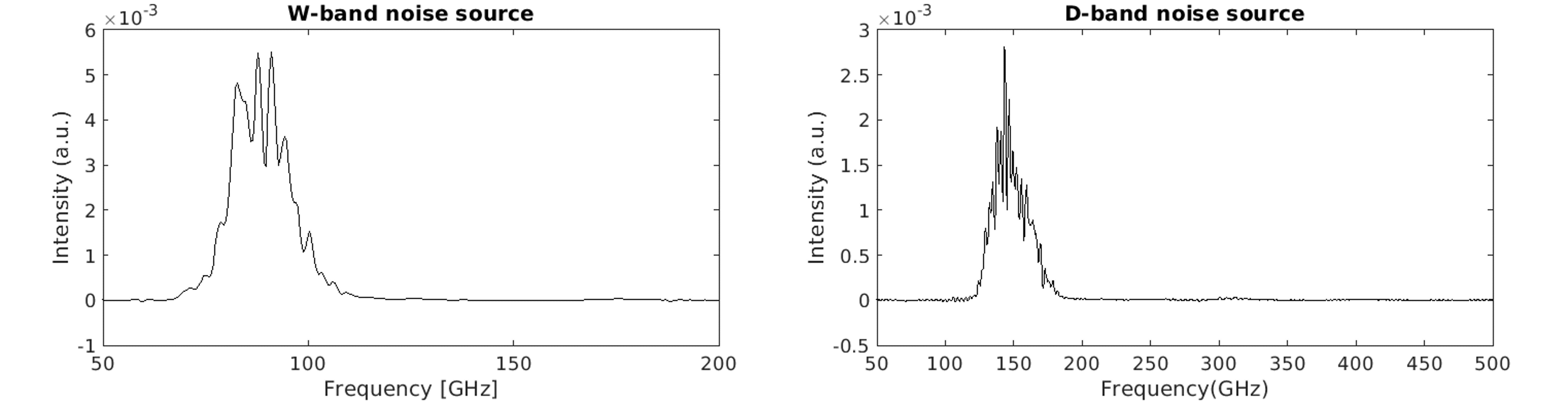}
\caption{\label{fig:spectra_tested_with_newFTS} Spectra of thermal noise source tested with a step-scan FTS. Left: spectrum of W-band thermal noise source. Right: spectrum of D-band thermal noise source. }
\end{figure}

The spectra of the two sources were tested with a Martin-Puplett type Fourier Transform Spectrometer (MP-type FTS for short) which is commissioned on the experimental advanced superconducting tokamak (EAST) \cite{Liu2016} to measure the electron cyclotron emission spectrum. The frequency resolution of the FTS is 2.7 GHz and the nominal frequency coverage ranging from 80 GHz to 350GHz.  The MP-type FTS designs incorporate four ports, with one input and one output port utilizing blackbody terminations, while the remaining input and output ports are dedicated to the source under test and the detector,respectively.  This FTS works in rapid continuous-scan mode. The spectra of the sources worked at amplified thermal noise mode (50$\Omega$ channel) and single-frequency mode (VCO channel) were tested, and the frequency step for single-frequency mode was 1 GHz. The VCO’s high-speed sweep mode is incompatible with the FTS, as the latter requires a time-invariant signal to acquire an interferogram. The integrated spectral output during a full sweep can thus be understood as the sequential emission of these validated discrete frequency components. The testing results are shown in figure \ref{fig:spectra_tested_at_EAST}. The spurious harmonics outside the designed bandwidth appear in all test results,   and this phenomenon is caused by the sampling errors \cite{Vidi2003,Peter2007} . Subsequent measurements with the EAST team's newly developed FTS  \cite{Meng2024} , which is operated in step-scan mode, showed no detectable harmonic components outside the target band (Figure \ref{fig:spectra_tested_with_newFTS}).

\subsection{Polarization Testing}\label{subsec3.4}
Since the source is linearly polarized, the polarization characteristics were tested in thermal noise mode at the maximum output power. 

\begin{figure}[H]
\centering
\includegraphics[width = 0.8\textwidth]{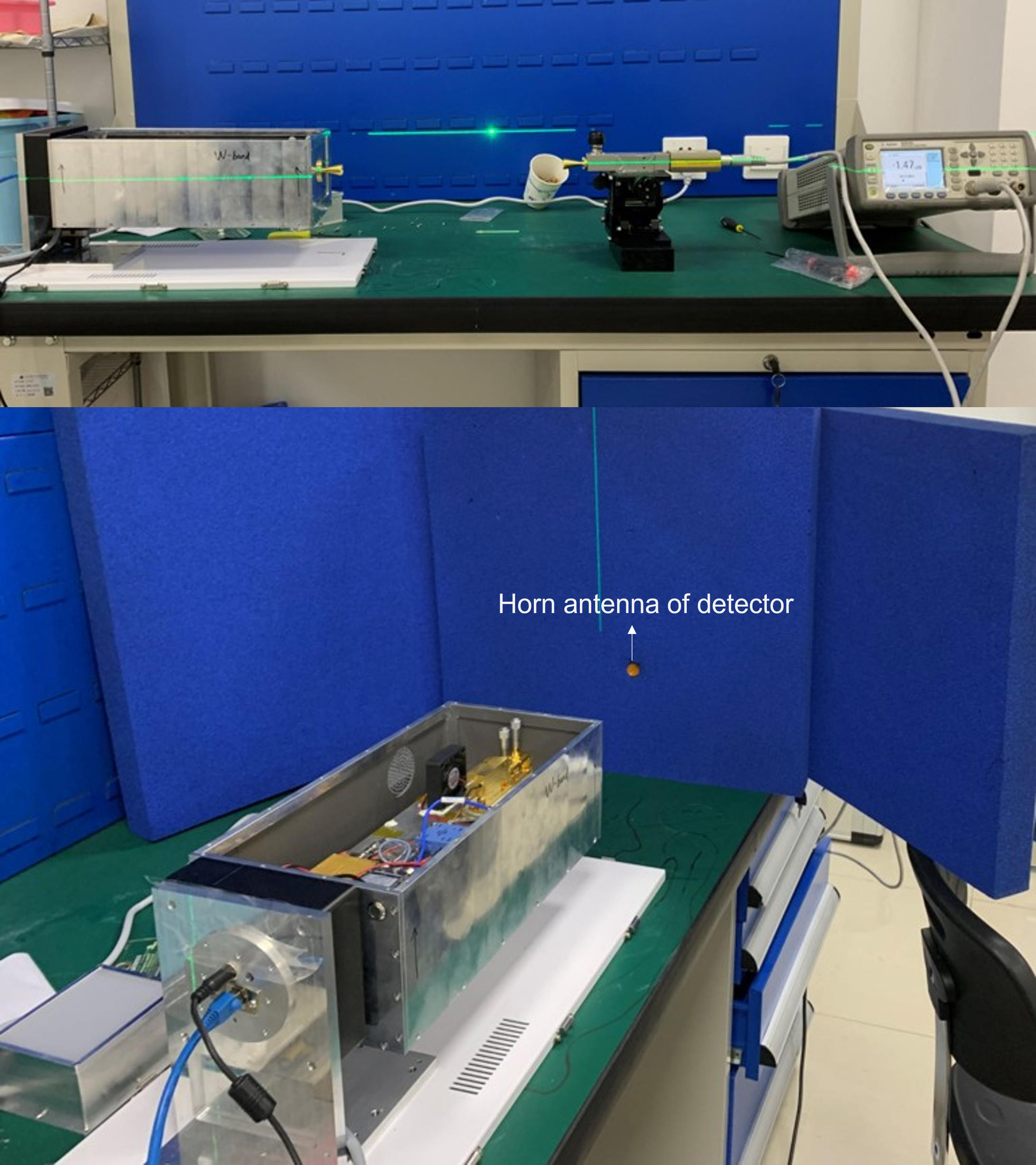}
\caption{\label{fig:polarizationTestSetup} Setup of W-band source polarization testing: A waveguide power sensor with a RCHO10R horn antenna was used to test the power of the transmitter, the horn distance between the source and the detector was larger than the far-field distance. The region surrounding the detector's horn was covered with  Dongshin  microwave  SAB-50 absorbers to reduce cross-polar leakage and reflection.  }
\end{figure}

For the W-band source polarization measurement, a free-space setup was employed. The transmitting source and the receiving detector were each connected to an RCHO10R horn antenna. The distance between the two antennas was 26 $cm$, which satisfies the theoretical far-field requiement of 16 $cm$ in theory. Alignment of the antennas was ensured using a cross-line laser. The received power was measured using a W8486A waveguide power sensor connected to the receiving antenna and a N1913A power meter. To minimize environmental reflections, Dongshin microwave SAB-50 absorbers were placed around the detector's horn. The testing setup are shown in Figure \ref{fig:polarizationTestSetup}.

For the D-band polarization measurement, a similar free-space configuration was used. The receiver consisted of an Eravant standard gain horn (model SAZ-2410-06-S1) connected in sequence to a Elmika WTT-03E/02E  WR10-to-WR-6.5 waveguide tapered transition, an Eravant SWC-101M-E1 W-band waveguide to 1.0 mm connector, and a U8489A power meter sensor. The distance between the transmitting and receiving horns was set to 40 cm, exceeding the calculated far-field distance of 36 cm based on the antenna apertures. Eccosorb HR-25 microwave absorbers were placed around the receiving antenna to suppress reflections.

The polarization characteristics were measured by recording the received power as the transmitter's rotation stage was turned through a range of angles. For each measurement, the rotation proceeded in a complete cycle (e.g., from 0\degree to 180\degree, then to –180\degree, and back to 0\degree for the W‑band, and from –180\degree to 180\degree and back to –180\degree for the D‑band). The rotation was not performed at uniform angular steps; instead, the step size was dynamically adjusted based on how close the measured power was to the expected minimum value, in order to capture the polarization null more accurately. At each angle, the power reading was recorded only after it had stabilized.

\begin{figure}[H]
\centering
\includegraphics[width = 0.8\textwidth]{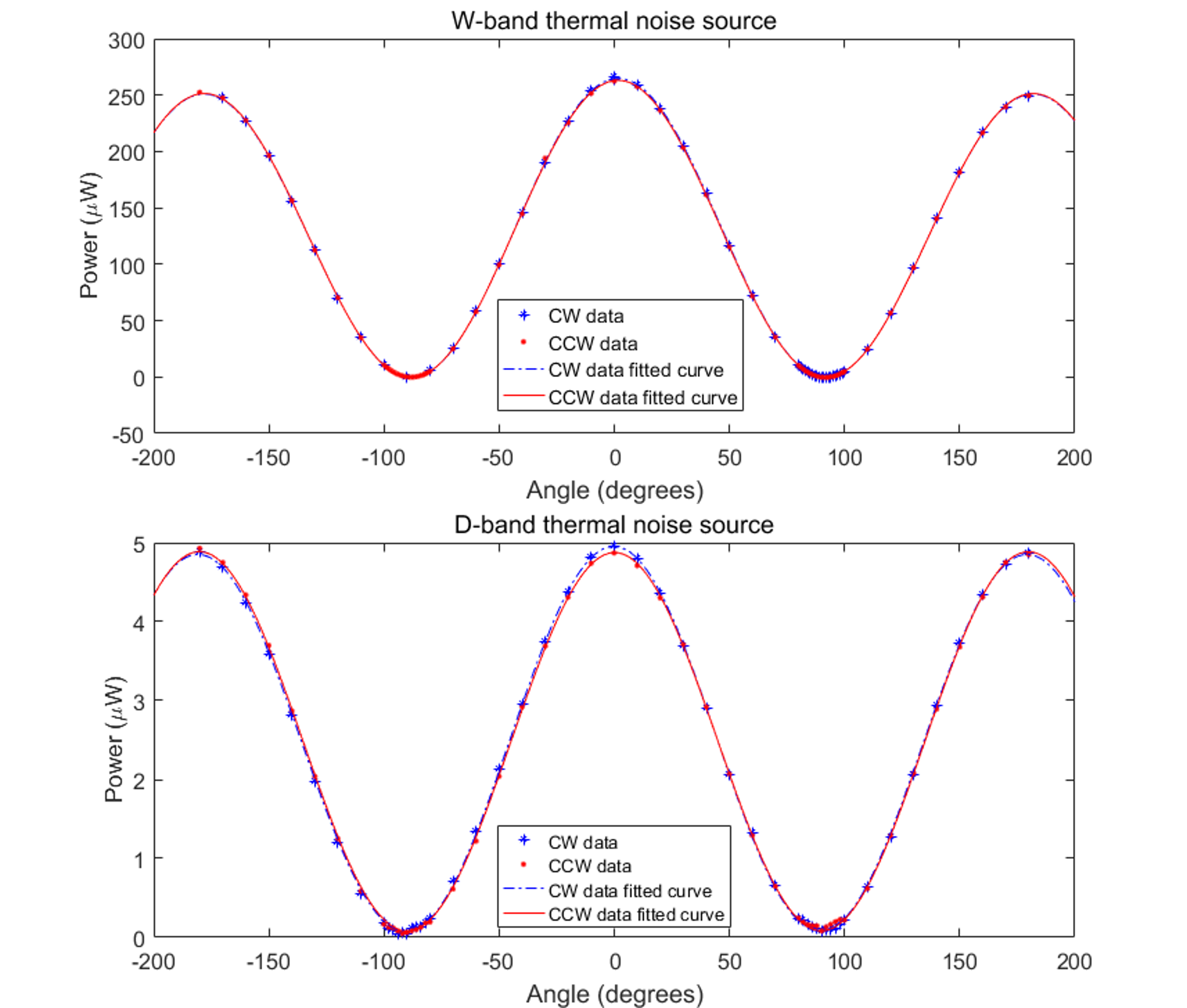}
\caption{\label{fig:polarizationtestdata} Polarization Responses of the W-band and D-band Sources. The received signal power is plotted as a function of the source rotation angle. In each subplot, blue asterisks and red dots denote the data points recorded during the clockwise (CW) and counterclockwise (CCW) rotational sweeps, respectively. The corresponding fitted curves for both sweep directions are also shown.}
\end{figure}

Figure\ref{fig:polarizationtestdata} shows the data, and the results exhibit the expected cosine pattern with a period of 180\degree, where peaks occur when the transmitter’s horn aligns with the detector’s horn, and troughs occur when they are perpendicular. However, for the results of W-band source, the peak at 0\degree is higher than those at $\pm 180\degree$, suggesting an additional modulation. This asymmetry can be explained by a misalignment between the rotational axis and the horn antenna’s position. As the source rotates, this misalignment causes the source to move radially, periodically reducing the alignment with the detector’s horn. This effect introduces a secondary modulation with a period of 360\degree, superimposed on the primary 180\degree signal. 
After accounting for this systematic effect, the following model was used to describe the modulation curve.:

\begin{equation}
  M(\theta) = C[\cos(2(\theta+\zeta))-\frac{\varepsilon+1}{\varepsilon-1}][D \cos(\theta-\gamma)+1] 
 % f(p,\theta) = p_0(1-p_1 cos^2({\theta}/{2}-p_2))(cos^2(\theta-p_3)+p_4 sin^2(\theta-p_3)),
\end{equation}

Where $\theta$ is the source polarization angle in degree, $\zeta$ is the polarization misalignment angle of the detector, representing the fixed angular offset between the detector' polarization axis and the source's reference frame.  $\varepsilon$ is cross-polar response ratio, C,D, and $\gamma$ are nuisance parameters to account for detector gain and the misalignment of the source rotation axis and detector horn. The measurement data from clockwise (CW) and counterclockwise (CCW) rotational sweeps were fitted separately. The resulting parameters are listed in Table \ref{tab:polarizationdatafitparameter}, where each entry is presented as the fitted value ± its $\pm 95\%$ confidence bounds.

\begin{table}[ht]
\caption{Fitting results for the polarization modulation model parameters in the W-band and D-band measurements.} 
\label{tab:polarizationdatafitparameter}
\begin{center}       
\begin{tabular}{|l|l|l|l|l|} %% this creates two columns
%% |l|l| to left justify each column entry
%% |c|c| to center each column entry
%% use of \rule[]{}{} below opens up each row
\hline
\rule[-1ex]{0pt}{3.5ex}  Parameter & W-band CW run & W-band CCW run & D-band CW run & D-band CCW run   \\
\hline
\rule[-1ex]{0pt}{3.5ex}  $C$ & $129.1 \pm 0.19$ & $128.8 \pm 0.22$ & $2.41 \pm0.01$ & $2.40 \pm 0.01$\\
\hline
\rule[-1ex]{0pt}{3.5ex}  $D$ & $0.027 \pm 0.0001$ & $0.023 \pm 0.002$ & $0.015 \pm 0.006$ & $-0.006 \pm 0.008$  \\
\hline
\rule[-1ex]{0pt}{3.5ex}  $\varepsilon$ & $-0.00114 \pm 0.0007$ & $-0.00105 \pm 0.0008$ & $0.0155 \pm 0.0023$ & $0.0166 \pm 0.0024$ \\
\hline
\rule[-1ex]{0pt}{3.5ex}  $\zeta$ & $-1.82 \pm 0.05$ & $-1.79 \pm 0.07$ & $0.49 \pm 0.17$ & $0.003 \pm 0.17$   \\
\hline
\rule[-1ex]{0pt}{3.5ex}  $\gamma$ & $-3.31 \pm 5.75$ & $-11.2 \pm 8.35$ & $44.9 \pm 23.9$ & $-79.1 \pm 39.3$  \\
\hline
\end{tabular}
\end{center}
\end{table} 

The measured cross-polarization response ratios differ significantly between the two bands:
approximately $-0.1\% $ for the W-band and $1.6\%$ for the D-band. The negative value for W-band source has no physical meaning and is not surprising in that since the expected cross-polar response value is so close to zero. The W-band system, configured with a matched pair of RCHO10R conical horns, achieved an excellent cross-polarization isolation of -30 dB ($\varepsilon \approx$ 0.001), demonstrating the high polarization purity of both the source and the identical antenna pair. Conversely, the D-band measurement yielded a cross-polarization ratio of -18 dB ($\varepsilon \approx$ 0.016). This difference is primarily attributed to the different receiver antenna used.  The D-band setup utilized a rectangular waveguide horn (SWC-101M-E1) at the receiver, whose cross-polarization performance is intrinsically lower than that of a conical horn. The conical horn (RCHO6R) used at the transmitter, similar in design to the W-band pair, is expected to exhibit superior polarization purity. Additionally, it should be noted that during the polarization sweep, the received power at the polarization null dropped to a level on the order of tens of nanowatts, which is below the specified sensitivity threshold (approximately -35 dBm, or $~0.3 \mu W$) of the power sensor used. This proximity to the detection limit may introduce additional uncertainty in the precise determination of the minimum power level, thereby affecting the accuracy of the extracted $\varepsilon$ value in the D-band measurement. Consequently, the measured $\varepsilon$ in the D-band is likely dominated by the combined effects of the polarization impurity of the receive antenna and the limitations near the sensor's detection threshold.

\section{Conclusion}\label{sec4}
Beam characterization is crucial for constraining the temperature to polarization leakage and the E-to-B leakage in polarization-sensitive CMB telescopes such as AliCPT-1. To this end, two rotating polarized sources in 90/150 $GHz$ band were designed and constructed to carry out the FFBM and FSLB calibrations. 

This paper has presented the comprehensive laboratory performance characterization of both calibration sources. The results demonstrate that all key parameters—including output power, dynamic range, long-term stability and spectra meet the design specifications and satisfy the stringent requirements for precision beam measurement.

The field performance of the D-band source has now been validated. During the first AliCPT-1 FFBM and FSLB campaign in May 2025, it operated robustly and confirmed that its output power and dynamic range in the field align with laboratory measurements and design expectations. Furthermore, the absence of scan-synchronous artifacts in the acquired beam maps provides direct evidence of its stability under deployment conditions. Both sources, including the fully validated 90 GHz unit, are on-site and ready for future campaigns.

In summary, we have developed and rigorously validated a pair of calibration sources for AliCPT-1. The D-band source has successfully transitioned from laboratory validation to field qualification, confirming its power, dynamic range, and stability. Together, these sources establish a reliable calibration foundation essential for achieving the telescope's scientific objectives. Detailed beam characterization results will be reported separately.

\section*{Disclosures}
The authors declare that there are no financial interests, commercial affiliations, or other potential conflicts of interest that could have influenced the objectivity of this research or the writing of this paper.

\section*{Code and Data Availability}
The data presented in this article are publicly available in FigShare 
at [10.6084/m9.figshare.29529119].

\section*{Acknowledgment}
This work was supported by National Key Research and Development Program of China (Grant No. 2020YFC2201604, 2022YFC2204900).

%%%%% References %%%%%

%\bibliography{report}   % bibliography data in report.bib
\bibliography{article}
\bibliographystyle{article}   % makes bibtex use spiejour.bst

%%%%% Biographies of authors %%%%%

%\vspace{2ex}\noindent\textbf{First Author} is an assistant professor at the University of Optical Engineering. He received his BS and MS degrees in physics from the University of Optics in 1985 and 1987, respectively, and his PhD degree in optics from the Institute of Technology in 1991.  He is the author of more than 50 journal papers and has written three book chapters. His current research interests include optical interconnects, holography, and optoelectronic systems. He is a member of SPIE.
\vspace{2ex}\noindent\textbf{Xufang Li} is a senior engineer at Institute of High Energy Physics, Chinese Academy of Sciences (IHEP, CAS). She earned her BS degrees in nuclear engineering and technology from East China University of Technology in 2009, her MS degree in nuclear technology and applications from  Beijing Normal University in 2012. Her work focuses on the calibration of astronomical telescopes, including the high energy telescope onboard \textit{Insight}-HXMT satellite and the AliCPT.

\vspace{2ex}\noindent\textbf{Congzhan Liu} is a senior scientist at Institute of High Energy Physics, Chinese Academy of Sciences (IHEP, CAS). He got his Doctor degree in Tsinghua University in 2006. He is dedicated to the design of astronomical instruments. In the last decade, he devoted himself to China's first CMB telescope -- AliCPT.

\vspace{2ex}\noindent\textbf{Aimei Zhang} is a senior engineer at IHEP, CAS. She received her BS degree in Materials Science and Engineering from the Inner Mongolia University of Technology in 2003, and MS degree in Mechanical Engineering from the Beijing Technology and Business University in 2006. Her  research lies in the mechanical design and analysis of space telescopes and ground-based Cosmic Microwave Background (CMB) telescopes. She is the chief structural designer for the Insight-HXMT satellite telescope, the 40G Hz CMB Telescope at Ali site, and the payload of the eXTP satellite.

\vspace{2ex}\noindent\textbf{Daikang Yan} conducted PhD research at Northwestern University and Advanced Photon Source during 2013 ~ 2019, and worked at the quantum sensors group at NIST, Boulder during 2019 ~ 2022. Her research is about superconducting detector development for various applications, and is currently developing superconducting transition-edge sensors for the AliCPT primordial gravitational wave project and the High Energy Photon Source at IHEP.

\vspace{1ex}\noindent Biographies and photographs of the other authors are not available.

%\listoffigures
%\listoftables

\end{spacing}
\end{document}